\pdfoutput=1
\documentclass[sigconf]{acmart}
\usepackage{color}
\usepackage{bbm}
\usepackage{multirow}
\usepackage[inline]{enumitem}
\usepackage{subfigure} 
\usepackage{sistyle}
\usepackage[ruled]{algorithm2e}
\usepackage{booktabs}
\usepackage{caption}
\SIthousandsep{,}
\usepackage{makecell}
\usepackage[skip=0pt]{caption}
\usepackage{amsmath}
\usepackage{hyperref}
\usepackage{array} 
\usepackage{graphicx}
\usepackage{xcolor}
\usepackage{acronym}
\usepackage[export]{adjustbox}

\newcommand{\heading}[1]{\vspace*{0.5mm}\noindent\textbf{#1.}}

\AtBeginDocument{%
  \providecommand\BibTeX{{%
    \normalfont B\kern-0.5em{\scshape i\kern-0.25em b}\kern-0.8em\TeX}}}

\makeatletter
\g@addto@macro\normalsize{%
  \abovedisplayskip 2pt plus1pt 
  \belowdisplayskip 2pt plus1pt
  \abovedisplayshortskip  2pt plus1pt%
  \belowdisplayshortskip  1pt plus1pt
}
\makeatother

\newcommand{\modelname}{LLM-QL\xspace}

\acrodef{IR}{information retrieval}
\acrodef{LLM}{large language model}

\AtBeginDocument{%
  \providecommand\BibTeX{{%
    Bib\TeX}}}

\copyrightyear{2026}
\acmYear{2026}
\setcopyright{cc}
\setcctype{by}
\acmConference[CIKM '26]{Proceedings of the 35th ACM International Conference on Information and Knowledge Management}{November 07--11, 2026}{Rome, Italy}
\acmBooktitle{Proceedings of the 35th ACM International Conference on Information and Knowledge Management (CIKM '26), November 07--11, 2026, Rome, Italy}
\acmDOI{10.1145/3799682.3839865}
\acmISBN{979-8-4007-2539-5/2026/11}
\makeatletter
\gdef\@copyrightpermission{
  \begin{minipage}{0.3\columnwidth}
   \href{https://creativecommons.org/licenses/by/4.0/}{\includegraphics[width=0.90\textwidth]{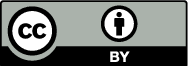}}
  \end{minipage}\hfill
  \begin{minipage}{0.7\columnwidth}
   \href{https://creativecommons.org/licenses/by/4.0/}{This work is licensed under a Creative Commons Attribution International 4.0 License.}
  \end{minipage}
  \vspace{5pt}
}

\author{Hengran Zhang}
\orcid{0009-0004-1144-1298}
\affiliation{
	\institution{State Key Laboratory of AI Safety}
	\institution{Institute of Computing Technology, Chinese Academy of Sciences}
	\institution{University of Chinese Academy of Sciences}
	\city{Beijing}
	\country{China}
}
\email{zhanghengran22z@ict.ac.cn}

\author{Keping Bi}
\orcid{0000-0001-5123-4999}
\authornote{Keping Bi is the corresponding author.}
\orcid{}
\affiliation{
	\institution{State Key Laboratory of AI Safety}
	\institution{Institute of Computing Technology, Chinese Academy of Sciences}
	\institution{University of Chinese Academy of Sciences}
	\city{Beijing}
	\country{China}
}
\email{bikeping@ict.ac.cn}

\author{Jiafeng Guo}
\orcid{0000-0002-9509-8674}
\affiliation{
	\institution{State Key Laboratory of AI Safety}
	\institution{Institute of Computing Technology, Chinese Academy of Sciences}
	\institution{University of Chinese Academy of Sciences}
	\city{Beijing}
	\country{China}
}
\email{guojiafeng@ict.ac.cn}

\author{Xiaojie Sun}
\affiliation{
	\institution{State Key Laboratory of AI Safety}
	\institution{Institute of Computing Technology, Chinese Academy of Sciences}
	\institution{University of Chinese Academy of Sciences}
	\city{Beijing}
	\country{China}
}
\email{sunxiaojie@ict.ac.cn}

\author{Shihao Liu}
\orcid{0009-0003-2736-3848}
\author{Daiting Shi}
\orcid{0000-0003-4926-3357}
\author{Dawei Yin}
\orcid{0000-0002-0684-6205}
\affiliation{
	\institution{Baidu Inc}
 \city{Beijing}
 \country{China}
}
\email{liushihao02@baidu.com}
\email{shidaiting01@baidu.com}
\email{yindawei@acm.org}

\author{Xueqi Cheng}
\orcid{0000-0002-5201-8195}
\affiliation{
		\institution{State Key Laboratory of AI Safety}
	\institution{Institute of Computing Technology, Chinese Academy of Sciences}
	\institution{University of Chinese Academy of Sciences}
	\city{Beijing}
	\country{China}
}
\email{cxq@ict.ac.cn}

\renewcommand{\shortauthors}{Hengran Zhang, Keping Bi, Jiafeng Guo, Xiaojie Sun, Shihao Liu, Daiting Shi, Dawei Yin, Xueqi Cheng}
\begin{document}

\title{Unleashing the Power of LLMs in Dense Retrieval with Query Likelihood Modeling}
\begin{abstract}
Dense retrieval is a fundamental task in Information Retrieval (IR), forming the foundation for downstream applications such as re-ranking and augmenting generation. Recently, large language models (LLMs) have exhibited remarkable semantic understanding capabilities, drawing significant interest from researchers in dense retrieval. While LLMs, as decoder-style generative models, excel in language generation, they often struggle to capture global information due to their lack of attention to subsequent tokens. Motivated by the classical word-based language modeling approach in IR, particularly the query likelihood (QL) model, we seek to exploit the generative advantages of LLMs through QL maximization. Instead of using QL estimation directly for document ranking, we introduce QL maximization as an auxiliary task to strengthen the backbone, thereby facilitating the subsequent contrastive learning of the retriever. 
We introduce our model, \modelname, which incorporates two key components: \textbf{Attention Block (AB)} and \textbf{Document Corruption (DC)}. AB blocks the attention of predictive tokens to the document tokens before the document's ending token, while DC corrupts a document by masking a portion of its tokens during prediction. Evaluations on the in-domain (MS MARCO) and out-of-domain dataset (BEIR) indicate \modelname's superiority over other LLM-based retrievers. Furthermore, comprehensive analyses also validate the efficacy of \modelname and its components. 
Our code can be found at \url{https://github.com/Trustworthy-Information-Access/LLM-QL}. 

\end{abstract}



\maketitle
\acresetall

\section{Introduction} 
Retrieval plays a crucial role in identifying relevant documents from a large-scale corpus in response to a query. 
Traditionally, retrieval emphasized lexical matching between query terms and passages, employing methods like BM25 \cite{lin2021pyserini} and the query likelihood (QL) model \cite{ponte2017language}. However, with the emergence of pre-trained language models (PLMs) such as BERT \cite{devlin2018bert}, representing passages or queries as dense vectors has become the dominant approach. These methods typically utilize two separate encoders to represent the query and passage, known as dual-encoder models.

Large language models (LLMs) are widely applied across fields \cite{kaddour2023challenges, xie2023adaptive, touvron2023llama} and increasingly explored for retrieval \cite{ma2024fine, jiang2023scaling, springer2024repetition}. Unlike bidirectional encoder-style PLMs (e.g., BERT), LLMs are typically decoder-style models with unidirectional attention. Next-token prediction on vast data strengthens their semantic understanding, but unidirectional attention can weaken global semantic representation, making them inferior to encoder PLMs in this regard. Despite their advanced semantics, applying LLMs to retrieval remains challenging. 
Recent studies have explored various approaches, such as repeating passages as input during encoding to ensure each token sees the original texts before and after it~\cite{springer2024repetition}, utilizing a bidirectional attention mechanism for encoding during relevance matching fine-tuning~\cite{behnamghader2024llm2vec}.

Since LLMs are decoder-based, it is natural to apply the classical query likelihood (QL) language modeling approach to IR. 
Recent work has explored QL-based ranking with encoder-decoder models \cite{zhuang2021deep}, and decoder-only models \cite{zhuang2023open}, which outperform word-based QL but substantially underperform BERT retrievers. 
This gap is attributed to generative models’ inability to capture multi-grade relevance and lack of contrastive learning. 
This raises the question: \textbf{do we have an effective way to leverage the generative capabilities of LLMs and unleash their potential in document representations for retrieval?}

To address these challenges, we propose \modelname, which seeks to harness the generative capabilities of LLMs for dense retrieval. Rather than modeling relevance matching through a generative process as seen in \cite{lesota2021modern, zhuang2021deep}, our approach retains discriminative modeling with a dual encoder and employs contrastive learning, while integrating query likelihood modeling as an auxiliary training task. This strategy allows the generalization ability of LLMs to be leveraged during the maximization of query likelihood, providing a stronger foundation for contrastive learning in relevance matching.

As we know, a potent encoder for dense retrieval should possess the ability to 1) effectively condense the semantics of documents or passages to a single vector, and 2) capture the potential query needs they may fulfill. We leverage query likelihood modeling to achieve the latter, and for the former capability, \modelname introduces two strategies to enhance semantic condensation: \textbf{Attention Block (AB)} and \textbf{Document Corruption (DC)}. Specifically, AB blocks the attention of a predictive token to the document tokens before the ending token during query token generation. This approach forces the document's semantics to be compressed into a single token for query prediction. 
Inspired by paragraph vectors with corruption \cite{chen2017efficient}, DC corrupts the document by randomly masking a portion of its tokens, aiming to condense as much document semantics as possible into the final representation and improve training efficiency.

We compare \modelname with various baselines using the widely-used MS MARCO \cite{bajaj2016ms} dataset as well as TREC DL 19 and 20 \cite{craswell2020overview}, focusing particularly on retrievers that are also based on LLMs. As our approach does not involve any large-scale pre-training, we emphasize comparisons with similar methods. Experimental results demonstrate that \modelname significantly outperforms LLM-based retriever baselines in terms of MRR and NDCG across different cutoffs. Zero-shot performance on the BEIR dataset \cite{thakur2021beir} also confirms the efficacy of \modelname. Additionally, we conducted extensive analysis on the model components, AB and DC, the effectiveness of \modelname based on different backbones, the ranking performance of QL estimation based on \modelname, efficiency, etc. Our study offers a feasible and promising direction for enhancing LLMs as retrievers in dense retrieval tasks.

\section{Related Work}

\begin{figure*}[t]
    \centering
    \includegraphics[width=\linewidth]{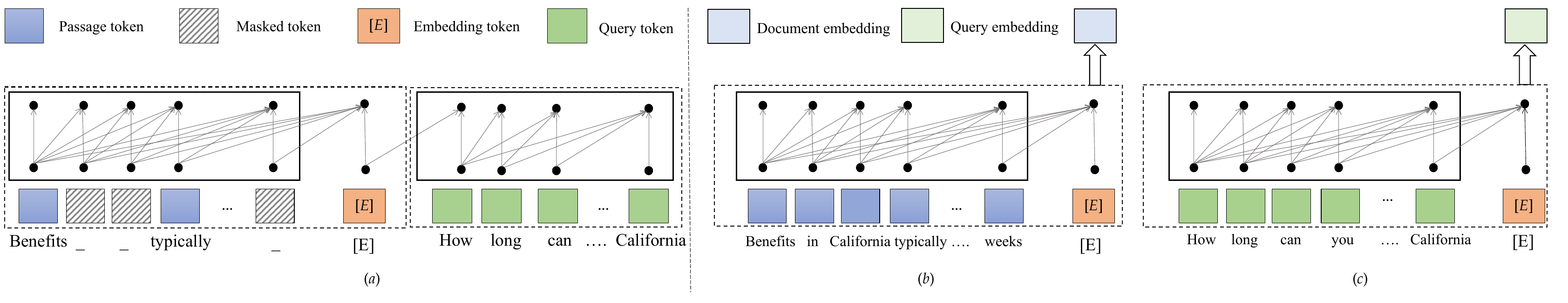}
    \caption{The details of \modelname. (a) QL learning: The input passage is moderately masked. When generating a query, passage tokens are blocked, and query generation solely focuses on the $[E]$. Contrastive learning contains (b) Document embedding and (c) Query embedding.  In our experiments, we use the special EOS token of LLMs as the $[E]$.}
    \label{fig:model}
\end{figure*}

Large language models (LLMs), e.g., LLaMa \cite{touvron2023llama, dubey2024llama}, Qwen \cite{bai2023qwen}, and GPT-4 \cite{achiam2023gpt},  have achieved excellent performance in various fields, especially in various subtasks of natural language generation (NLG). 
Recently, many studies \cite{ma2024fine, li2024llama2vec, behnamghader2024llm2vec, lee2024nv, springer2024repetition, moreira2024nv} have applied decoder-only architecture LLMs to dense retrieval to encode query and document embeddings as representation vectors. 
For example, 
RepLLaMA \cite{ma2024fine} directly replaces the PLMs with LLMs to encode the representation of query and document, and empirical evidence shows that LLM-based retrievers trained using a simple fine-tuning strategy outperform LM-based retrievers trained using complex training strategies. 
LLM2VEC \cite{behnamghader2024llm2vec, lee2024nv} replaced the decoder-only unidirectional attention with bidirectional attention, further improving retrieval performance. 
Echo \cite{springer2024repetition} repeated the input twice in context and extracted embeddings from the second occurrence to address an architectural limitation of autoregressive models. 
Unlike our strong baselines (like RepLLaMA, Echo, and LLM2VEC), we propose a novel approach that leverages the generative strengths of LLMs through QL maximization to enhance the backbone for subsequent contrastive learning of the retriever. 

\section{QL-based LLM Retriever (LLM-QL) } 


To leverage LLMs' generation capability to better serve retrieval, we propose \modelname, which incorporates QL modeling before fine-tuning LLMs with regular contrastive learning. The QL maximization objective activates LLMs' proficiency in language modeling for query generation, serving as a bridge between their powerful generative abilities and relevance understanding. 
To effectively condense the semantics of documents or passages into single vectors, we propose two strategies in \modelname: \textbf{Attention Block (AB)} and \textbf{Document Corruption (DC)}. 
Overall, \modelname employs a two-stage training process: query likelihood (QL) modeling followed by contrastive fine-tuning, as illustrated in Figure \ref{fig:model}. 
\begin{figure}[t]
    \centering
   \includegraphics[width=0.4\textwidth]{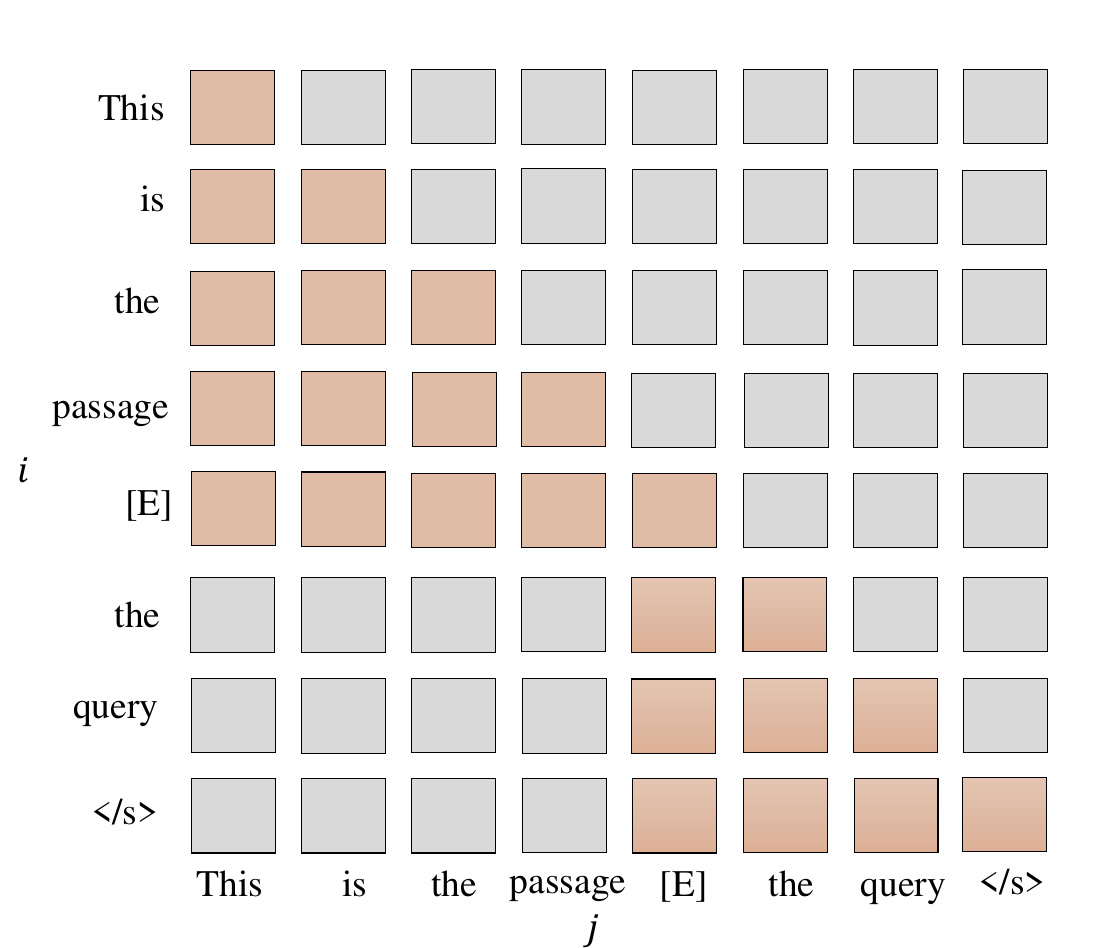}
    \caption{The attention matrix during query generation. The gray-shaded regions indicate that when generating token $i$, token $j$ cannot be attended to.} 
    \label{fig:mask_attention}
\end{figure}

\subsection{QL Modeling} 
Given a query and its annotated relevant documents, the first training stage is QL modeling, which involves two strategies, i.e., \textit{Attention Block (AB)} and \textit{Document Corruption (DC)}.

\heading{Attention Block (AB)} 
To compress a document's semantics into a single token for query generation, AB employs a constrained attention mechanism during query generation. 
As illustrated in Figure \ref{fig:mask_attention}, this attention masking creates an information bottleneck, forcing the model to generate queries exclusively based on the semantics captured by the document-ending token.

\heading{Document Corruption (DC)} 
\label{sec:heading:dc}
To condense more global semantics of the document into a single vector, we corrupt the document by stochastically masking its tokens. 
DC randomized masking compels the model to reconstruct queries from incomplete inputs. This design not only enhances robustness but also encourages the model to consolidate scattered semantic clues into the [E] vector, rather than relying on specific keywords.


\subsection{Contrastive Learning} 

 
The LLM-based retriever is then optimized via contrastive learning with the following objective function: 
\begin{equation}
    \mathcal{L}_{CL}(q, d^+, D^-)=-log\frac{exp(s(q, d^+))}{\sum_{d\in \{d^+\} \cup D^-}exp(s(q, d))},
\end{equation}
where $d^+$ is a relevant document and $D^-$ represents a collection of negative documents for the query $q$, usually including hard negatives and in-batch negatives \cite{karpukhin2020dense}.

\section{Experimental Setup}
\subsection{Datasets} 
We conduct experiments using the MS MARCO passage retrieval dataset  \cite{bajaj2016ms}, containing about 8.8 million passages. 
For in-domain evaluation, we employ the MS MARCO dev set (6,980 queries), TREC DL19 (43 queries)~\cite{craswell2020overview}, and TREC DL20 (54 queries) \cite{craswell2021overview}.  We further assess zero-shot generalization on 14 datasets that are publicly available in BEIR \cite{thakur2021beir} to evaluate the models. 

\subsection{Training Settings}
The proposed \modelname is implemented on the LLaMA-2-7B base through two training processes using the same queries: i.e., query likelihood modeling and contrastive learning.  
The training is on a machine with 8× Nvidia A800 (80GB) GPUs using Zero Redundancy Optimizer (ZeRO-3) from DeepSpeed \cite{rasley2020deepspeed}.  
For the  \textbf{QL modeling}, we perform 2 epochs on MS MARCO q-d pairs in total, with a batch size of 512, query max length of 200, passage max length of 200, and a learning rate of 1e-5. 
The default masking ratio of the passage is 0.6.  
For the \textbf{contrastive learning}, following the training details of our relevant baseline \cite{ma2024fine}, the model is trained for 1 epoch, batch-size 32, learning rate 1e-4, and gradient accumulation steps 4.

\begin{table}[t]
\small
  \centering
    \renewcommand{\arraystretch}{0.9}
   \caption{Passage retrieval performance (\%) on MS MARCO dev. 
   $^\star$ means that we reproduce the models based on Llama2-7B. 
     \textbf{Bold} and \underline{underlined} represent the best and second best performance, respectively. 
      `$^{\dagger}$' indicates significant improvement over the best-performing baseline using the paired t-test ($p$ < 0.05).}
   \setlength\tabcolsep{1.5pt}
    \begin{tabular}{ll ll}
    \toprule
 Method & MRR@10 & Recall@1000 \\ \hline
    RepLLaMA (LAFT) \cite{ma2024fine}  & 41.2  & \textbf{99.4}   \\   
      RepLLaMA (FFT) \cite{ma2024fine}  & 41.6  & -  \\  
     Summarize$^\star$(LA) \cite{jiang2023scaling}  & 41.0  & \textbf{99.4}\\   
    Echo$^\star$ (LAFT) \cite{springer2024repetition}  & 41.5  & \textbf{99.4} \\  
   LLM2VEC$^\star$ (LAFT+LAFT) \cite{behnamghader2024llm2vec} & 41.8 & 99.3 \\
     LLM2VEC$^\star$ (FFT+LAFT) \cite{behnamghader2024llm2vec}  &  \underline{41.9} &  \underline{99.3} \\  
     \midrule
    \modelname (LA+LAFT) & \underline{41.9}   & \textbf{99.4} \\
  \modelname (FFT+LAFT)  & \textbf{42.4} \rlap{$^{\dagger}$}   & \textbf{99.4}  \\
        \bottomrule
    \end{tabular}
   
  \label{tab:passage_retrieval}%
\end{table}%


\begin{table*}[t]
  \centering
    \small
   \renewcommand{\arraystretch}{0.9}
  \caption{Retrieval performance (\%) upon different LLMs.  `$^{\dagger}$' indicates significant improvements over LLM2VEC  ($p$<0.05, paired t-test). `$^{\ddagger}$' indicates significant improvements over LLM2VEC  on Fisher's test ($p$<0.05). \textbf{Bold} represents the best performance among the same group. ``Q'', ``L'' mean ``Qwen'' and ``Llama'', respectively. ``N'', ``R'' mean ``NDCG'' and ``Recall'', respectively.}
    \begin{tabular}{cllllllllcll}
    \toprule
    \multirow{3}[5]{*}{LLM} & \multicolumn{1}{l}{\multirow{3}[5]{*}{Method}} & \multicolumn{6}{c}{Dev}                       & \multicolumn{2}{c}{DL19} & \multicolumn{2}{c}{DL20} \\
\cmidrule(r){3-8}  \cmidrule(r){9-10}   \cmidrule(r){11-12}          &       & \multicolumn{2}{c}{MRR} & \multicolumn{2}{c}{NDCG} & \multicolumn{2}{c}{Recall} & \multicolumn{1}{c}{NDCG} & \multicolumn{1}{c}{Recall}  & \multicolumn{1}{c}{NDCG} & \multicolumn{1}{c}{Recall} \\
\cmidrule(r){3-4}   \cmidrule(r){5-6} \cmidrule(r){7-8} \cmidrule(r){9-9} \cmidrule(r){10-10} \cmidrule(r){11-11} \cmidrule(r){12-12}            &       & @10 & @20 & @10 & @20 & @100 & @1000  & @10 & @100  & @10  & @100\\
    \midrule
    \multirow{2}[1]{*}{Qwen3-0.6B} 
    & LLM2VEC & 38.7  & 39.4  &45.6 & 48.1 & 91.9 & 98.6  & \textbf{72.6} & 51.4     & \textbf{68.0} & 54.6 \\
    & LLM-QL & \textbf{39.5}$^{\dagger}$ & \textbf{40.2}$^{\dagger}$ & \textbf{46.3}$^{\dagger}$ & \textbf{48.7}$^{\dagger}$ & \textbf{92.5}$^{\dagger}$ & \textbf{98.9}$^{\dagger}$  & 72.5 & \textbf{52.2}  & 67.7 & \textbf{55.5}$^{\ddagger}$  \\
    \midrule
    \multirow{2}[1]{*}{Qwen2.5-7B} 
    & LLM2VEC & 41.9  & 42.6  & 48.8 & 51.3 & 94.3 & 99.4  & 73.5 & 56.2   & 72.0  & 60.8  \\
    & LLM-QL & \textbf{42.4}  & \textbf{43.1}  & \textbf{49.3}$^{\dagger}$ & \textbf{51.8}$^{\dagger}$ & \textbf{94.7}$^{\dagger}$ & \textbf{99.5}   & \textbf{73.8} & \textbf{56.6}  & \textbf{72.9}$^{\ddagger}$  & \textbf{61.3}$^{\ddagger}$  \\
    \midrule
    \multirow{2}[1]{*}{Llama2-7B} 
    & LLM2VEC &  41.9 &  42.5  & 48.8  &  51.2  & 94.2  & 99.3  & 72.9  & 54.8  & \textbf{74.4}  & 61.0\\
    & LLM-QL & \textbf{42.4$^{\dagger}$} & \textbf{43.1$^{\dagger}$} & \textbf{49.2} & \textbf{51.7$^{\dagger}$} & \textbf{94.6$^{\dagger}$} & \textbf{99.4} & \textbf{73.6}  & \textbf{55.4} & 72.4  & \textbf{61.2}$^{\ddagger}$ \\
    \bottomrule
    \end{tabular}%
  \label{tab:different_llm}%
\end{table*}%

\subsection{Baselines} 
    

The following LLM-based retrievers are the closest baselines: 
\begin{itemize}[leftmargin=*,itemsep=0pt,topsep=0pt,parsep=0pt]
    \item \textbf{RepLLaMA:} It extracts the final layer hidden state of the last token as the dense embedding for the query or the passage. 
    
    \item \textbf{Echo embedding (Echo):} 
        It repeats the query or passage in context and extracts the last token embedding.
       
    \item \textbf{Summarize embedding (Summarize):} 
    It compresses the input text into an embedding using the prompt: \textsl{This sentence: $q/d$ means in one word:}``, and exact the last token embedding. 
    \item \textbf{LLM2VEC:} 
    LLM2VEC directly replaces the causal attention of decoder-only LLMs with bidirectional attention. 
    Furthermore, LLM2VEC introduces the masked next token prediction task training to enable LLMs to utilize bidirectional attention. 
\end{itemize} 
For a fair comparison, we employ Llama2-7B \cite{touvron2023llama} as the LLM backbone for all baselines. 
Both LLM2VEC and our model have two-stage training. 
We use full-parameter fine-tuning (\textbf{FFT}) for the first training stage and LoRA fine-tuning (\textbf{LAFT}) for the second one. 

\section{Experimental Results}
\subsection{In-domain Retrieval Performance}

\begin{table*}[t]
  \centering
  \small
  \caption{Zero-shot NDCG@10 performance (\%) on BEIR benchmark. \textbf{Bold} and \underline{underlined} represent the best and second best performance, respectively. The backbone of the LLM-based retriever is the Llama2-7B.}
    \renewcommand{\arraystretch}{0.9}
   \setlength\tabcolsep{3.5pt}
    \begin{tabular}{lccccccccccc}
    \toprule
    Method & BM25  & BRET  & GTR-XXL & CPT-XL & Ada-2 & SGPT  & RepLLaMA & LLM2VEC(FFT) & LLM2VEC(LAFT) & \modelname(FFT) & \modelname(LAFT) \\
    size  &       & 110M  & 4.8B  & 175B  &       & 5.8B  & 7B    & 7B    & 7B    & 7B    & 7B \\
    \midrule
    DBPedia & 31.8  & 31.4  & 40.8  & 43.2  & 40.2  & 39.9  & 43.7  & 45.1 & \underline{45.1}  & 44.0  & \textbf{45.2} \\
    FiQA  & 23.6  & 25.2  & \underline{46.7}  & \textbf{51.2} & 41.1  & 37.2  & 45.8  & 44.8 &39.6 & 44.7  & 46.2  \\
    NQ    & 30.6  & 46.7  & 56.8  & -    & 48.2  & 52.4  & 62.4  & 62.4  & 62.8  & 63.1  & \textbf{64.9} \\
    HotpotQA & 63.3  & 48.8  & 59.9  & 68.8  & 65.4  & 59.3  & 68.5  & 69.1 & 69.3 & 67.8  & \textbf{69.4} \\
    NFCorpus & 32.2  & 26.0  & 34.2  & \textbf{40.7} & 35.8  & 36.2  & 37.8  & 36.2 & 33.7  & 36.6  & \underline{36.8}  \\
    ArguAna & 39.7  & 26.5  & \underline{54.0}  & 43.5  & \textbf{56.7} & 51.4  & 48.6  & 40.8  & 40.8  & 40.5  & 40.2  \\
    C-FEVER & 16.5  & 18.7  & 26.7  & 22.3  & 23.7  & 30.5  & \textbf{31.0} & 30.3  & 25.3  & 24.9  & \underline{30.4}  \\
    FEVER & 65.1  & 68.2  & 74.0  & 77.5  & 77.3  & 78.3  & \textbf{83.4} & 78.0  & 70.0  & 76.8  & \underline{82.8}  \\
    T-COVID & 59.5  & 61.5  & 50.1  & 64.9  & 81.3  & \textbf{87.3}  & 84.7  & \underline{84.8}  & 80.8  & 83.7  & 84.0  \\
    Touche & \textbf{44.2} & 25.9  & 25.6  & 29.1  & 28.0  & 25.4  & 30.5  & 39.1 & 32.3  & 32.1  & \underline{37.0}  \\
    CQA   & 32.5  & 28.2  & 39.9  & -     & 41.7  & 38.1  & 37.4  & 41.3  & 37.0  & 39.4  & \textbf{41.8} \\

    Quora & 78.9  & 78.7  & \textbf{89.2} & 63.8  & 87.6  & 84.6  & 86.8  & 87.7  & 88.4  & \underline{88.6}  & 87.6  \\
    SCIDOCS & 14.1  & 11.3  & 16.1  & -    & 18.6  & \textbf{19.7} & 18.1  & 17.7  & 17.6  & 18.9  & \underline{19.0}  \\
    SciFact & 67.9  & 53.3  & 66.2  & 75.4  & 73.6  & 74.7  & \underline{75.6}  & \textbf{76.6} & 73.2 & 75.5  & 73.6  \\
    \midrule
     Avg & 42.9  & 39.3  & 48.6  &   -    & 51.4  & 51.1  & \underline{53.9}  & 53.8  & 51.1  & 52.6  & \textbf{54.2} \\
    \bottomrule
    \end{tabular}%
  \label{tab:beir}%
\end{table*}%


Experimental results are reported in Tables~\ref{tab:passage_retrieval}\&\ref{tab:different_llm}. 
From these two tables, we have the following observations:


\heading{Uni-directional Attention vs Bi-directional Attention} 
While experiments demonstrate that bidirectional attention mechanisms generally outperform unidirectional ones in capturing global semantics (as shown by the LLM2VEC versus RepLLaMA comparison in Table \ref{tab:passage_retrieval}), our proposed \modelname fundamentally challenges this paradigm. 
By integrating query likelihood modeling into decoder-only LLMs, even with an unidirectional attention mechanism, document semantics can be learned well in the vector representation for relevance matching. 
This finding reveals that QL maximization improves global semantic understanding, thereby helping unleash the potential of LLMs in dense retrieval tasks. 
\heading{\modelname versus LLM-based Retriever Baselines} 
Both LLM2VEC and our \modelname employ two-stage training approaches. During the first-stage training phase, we implemented both full-parameter fine-tuning (FFT)  and LoRA fine-tuning (LA) for both models. 
The FFT approach allows more capacity for \modelname to learn query generation than LoRA fine-tuning and demonstrates better effectiveness. 
Notably, it attains an \textit{MRR@10 of 0.424} on the MS MARCO-dev set, significantly outperforming the baseline methods and achieving a new state-of-the-art performance on this large-scale dataset. 
For the TREC DL, the small-scale datasets,  \modelname's performance in the top ranks (NDCG@10) on TREC DL is not state-of-the-art, likely attributable to greater statistical variability inherent in the limited data size. Conversely, \modelname achieves consistently superior results in deeper recall (Recall@100 and Recall@1000), demonstrating a stable advantage across both TREC DL 2019 and 2020.

\subsection{Zero-shot Out-of-domain Performance} 
We also examine the generalization capabilities of retrievers trained with MS MARCO on the BEIR benchmark \cite{thakur2021beir}, as shown in Table \ref{tab:beir}. 
We can observe that: 
\begin{enumerate*}[leftmargin=*,itemsep=0pt,topsep=0pt,parsep=0pt]
\item LLMs-based retrievers outperform BM25 and BERT-based retrievers on BEIR, demonstrating the powerful generalization capabilities inherent in large language models. 
\item Demonstrating strong retrieval capabilities, \modelname ranks first or second on 9 tasks and achieves state-of-the-art performance on the BEIR benchmark (avg. NDCG@10 = \textbf{0.542}), with particularly excelling on search-oriented tasks like NQ and HotpotQA. 

\end{enumerate*}

\begin{figure}
    \centering
    \includegraphics[width=\linewidth]{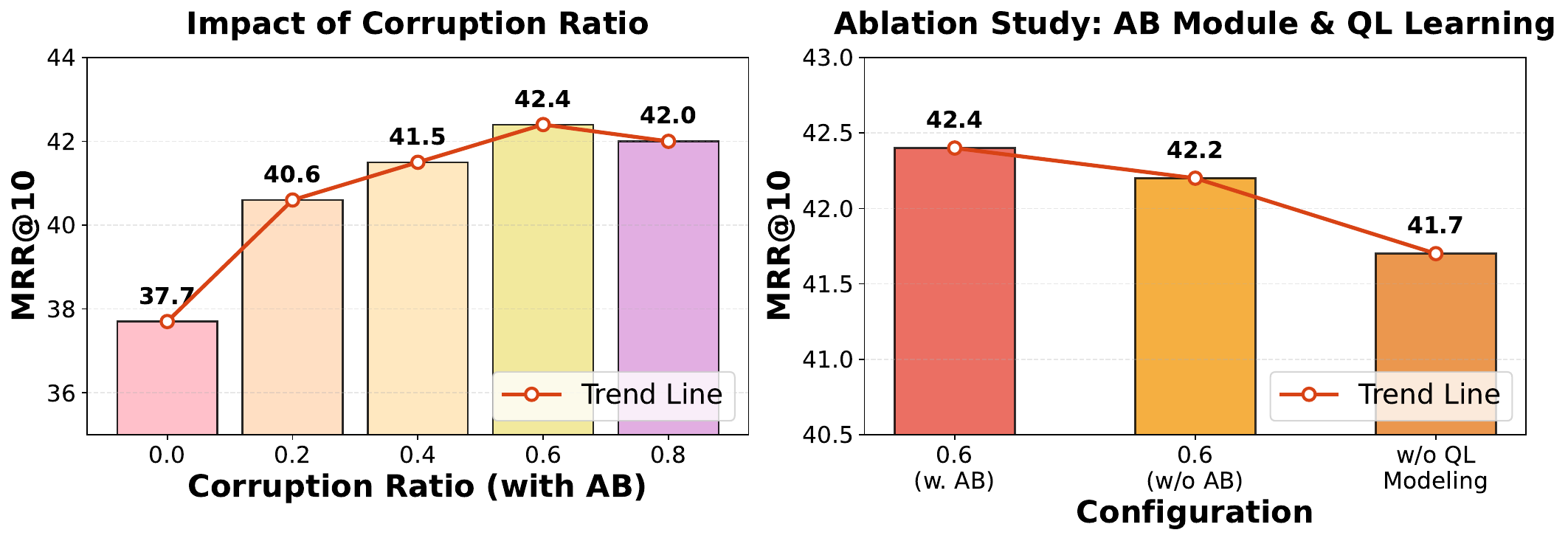}
    \caption{Ablation studies on AB, DC, and the QL learning on the MS MARCO-dev}
    \label{fig:Ablation}
\end{figure}

\subsection{Ablation Study} 
Figure \ref{fig:Ablation} presents the ablation study of attention block (AB), and document corruption (DC).

\heading{Attention Block and Document Corruption} 
We found that removing AB  or DC has an impact on the performance of the model. 
However, the importance of the two for performance gains is not the same. 
Removing DC has a greater impact on performance than removing AS, which illustrates the importance of the corruption mechanism.

\heading{Corruption Ratio}
A small corruption ratio may not be sufficient to assist LLMs in better condensing the semantics of passages. 
Conversely, an excessively large corruption ratio can hinder the model from comprehending the input passage and compress it into a query, as it becomes challenging for the model to process.

\subsection{Variants of Embedding Tokens $[E]$} 
Besides the current single-vector representation using a single end-of-sentence (EOS) token ``</s>'' as the query or document embedding token $[E]$, we also investigate multi-vector presentations: 
1) Replicating EOS tokens, i.e., ``</s></s></s></s>'', denoted as ``</s>$*4$'';
2) Multiple different tokens, using four newly defined tokens ``<s1><s2><s3><s4>'', denoted as ``s1234''. 
The final representation is obtained through mean pooling over the embeddings of the multiple tokens. 
The performance of the three results after fine-tuning is shown in Table \ref{tab:e}. 
We observe that:
\begin{enumerate*}
[leftmargin=*,itemsep=0pt,topsep=0pt,parsep=0pt]
\item Multi-vector representations can result in better recall but worse MRR and NDCG than single-vector representations; 
\item Representations with multiple replicated EOS tokens perform better than using different new tokens, which may be due to the insufficient training of the newly defined tokens. 
\end{enumerate*}

\begin{table}[t]
\centering
 \caption{Investigating variants of $[E]$ on MS MARCO dev. ``</s>*4'' means ``</s></s></s></s>''. ``s1234'' denotes <s1><s2><s3><s4>. \textbf{Bold} represents the best performance.}
   \setlength\tabcolsep{2.5pt}
    \begin{tabular}{l cccc}
    \toprule
 $[E]$ & MRR@10 & NDCG@10 & Recall@100 & Recall@1000 \\ 
 \midrule
 </s> (\modelname) & \textbf{42.4} & \textbf{49.2} & \textbf{94.6} & 99.4 \\
 \midrule
 </s>*4 & 42.1  & 49.0  & 94.5  & \textbf{99.5} \\
 s1234 & 42.0  & 48.9 & \textbf{94.6}  & 99.4 \\
\bottomrule
    \end{tabular}
  \label{tab:e}%
\end{table}%

\begin{table}[t]
  \centering
   \caption{Performance (\%) of retrieval and re-ranking task using query likelihood model on MS MARCO ranking dataset. \textbf{Bold} represents the best performance. The superscript $^{\dagger}$ indicates significant improvement over the best-performing baseline using the paired t-test ($p$ < 0.05).}
  \small
   \renewcommand{\arraystretch}{0.8}
   \setlength\tabcolsep{2pt}
    \begin{tabular}{l cccc c}
    \toprule
Method&  NDCG@1 & NDCG@3 & NDCG@10 & MRR@10 \\ 
\midrule

BM25  & 10.4	& 17.4 & 23.4	& 18.7 \\ 
QLM-JM & \phantom{1}9.6 &	15.9	& 21.8 &	17.4 \\
QLM-D & \phantom{1}8.3 & 13.7 &	18.7 & 14.9 \\

\midrule

BM25 + QLM-T5 & 17.8 & 28.2 & 36.5  &  30.0  \\

BM25  + Llama2-7B & \phantom{1}1.0  &  \phantom{1}1.7 & \phantom{1}2.9 &  \phantom{1}2.1  \\  
\midrule
\multicolumn{5}{l}{BM25  + \modelname (w/o contrastive learning)} \\
Inference   w/ AB &  16.0  & 26.0 & 33.7 & 27.6  \\
Inference  w/o AB &  17.1 &   27.1 &   34.8 & 28.7 \\
\midrule
BM25  + \modelname & \textbf{22.3}$^{\dagger}$ & \textbf{33.4}$^{\dagger}$ & \textbf{41.0}$^{\dagger}$ & \textbf{34.7}$^{\dagger}$ \\
\bottomrule
    \end{tabular}
  \label{tab:qlm}%
\end{table}%

\section{Conclusion and Future Work} 
In this work, we propose \modelname inspired by the query likelihood modeling, which aims to utilize generation capabilities for dense retrieval.  
\modelname contains two training stages: QL learning and Contrastive Learning. 
For QL learning, in order to sufficiently condense the semantics of documents or passages to a single vector, we propose two strategies: Attention Block (AB) and Document Corruption (DC).
Through this simple step of adaptation, our method remarkably enhances the model’s fine-tuned performance on dense retrieval benchmarks.  
We also conduct comprehensive analyses of the
components of LLMs, other options of the model, and the ranking
performance of query likelihood modeling based on \modelname. 

For future work, we would like to focus on two promising directions: 
1) Large-scale pre-training paradigm: While our method demonstrates strong performance with standard training data, scaling up through pseudo query generation warrants systematic exploration. 
2) Efficiency of LLM-based dense retrievers: 
Future research can focus on enhancing the search efficiency of LLM-based dense retrievers through dimensionality reduction techniques.


\begin{acks}
This work was funded by New Generation Artificial Intelligence-National Science and Technology Major Project of No. 2025ZD0123301, the National Natural Science Foundation of China (NSFC) under Grants No. 62302486 and No. 62441229, the Innovation Project of ICT CAS under Grants No. E561010 and E361140, and the Strategic Priority Research Program of the CAS under Grants No. XDB0680102. 
\end{acks}

\section*{GenAI Usage Disclosure}
During the preparation of this work, the author(s) only used ChatGPT (GPT-4o) or DeepSeek for writing and language polishing to improve the clarity, grammar, and readability of selected paragraphs. 
No other Generative AI tools were used in the research (including experimental design, data generation, or result interpretation) or in the writing of this paper. The author(s) remain solely responsible for all scientific and ethical aspects of this work. 


\bibliographystyle{ACM-Reference-Format}
\balance
\bibliography{reference}
\end{document}